\documentclass[12pt]{article} 

\newcommand{\be}{\begin{eqnarray}}
\newcommand{\ee}{\end{eqnarray}}
\newcommand{\ave}[1]{\left\langle #1 \right\rangle}

\usepackage{amsmath}
\usepackage{epsfig} 
\usepackage{amssymb}
\usepackage{graphicx}
\usepackage{subfigure}
\usepackage{cite}
\usepackage{alltt}

\begin{document}

\begin{titlepage}

\begin{flushright}
\end{flushright}
\vspace{1mm}
\begin{center}{\bf\Large SHAREv2: fluctuations and a comprehensive treatment\\ of decay feed-down $\dag$}
\end{center}

\vspace{3 mm}

\begin{center}
{\large  G. Torrieri$^{\,a}$,}
{\large  S. Jeon$^{\,a,b}$,}
{\large  J. Letessier$^{\,c,d}$,} and
{\large  J. Rafelski$^{\,d}$}  
\\
\vspace{0.5cm} 
{\em $^a$ Department of Physics, McGill University, Montreal, QC H3A-2T8, Canada}\\ 
{\em $^b$ RIKEN-BNL Research Center, Upton NY, 11973, USA}\\
{\em $^c$ Laboratoire de Physique
Th\'eorique et Hautes Energies {$^{\ddag}$}\\ Universit\'e Paris 7, 2 place
Jussieu, F--75251 Cedex 05, France}\\
{\em $^d$ Department of Physics, University of Arizona,
Tucson, AZ~85721, USA}
\end{center}
\vspace{1mm}
\begin{abstract}
This the user's manual for SHARE version 2. SHARE  \cite{share_orig} (Statistical
Hadronization with Resonances) is
 a collection of programs designed for the statistical analysis of
particle production in relativistic heavy-ion collisions.
While the structure of the program remains similar to v1.x, v2 provides several 
new features  such as evaluation of statistical fluctuations of particle yields,
 and a greater   versatility, in particular regarding
decay feed-down and input/output structure.   This article describes all 
the new features, with emphasis on statistical fluctuations.
\end{abstract}

\vfill
\noindent \textbf{Updates available from:}
{\tt http://www.physics.arizona.edu/$\tilde{\phantom{~}}$torrieri/SHARE/share.html }
and the authors upon request.
\vskip 0.4cm
\footnoterule
\noindent
{\footnotesize
\begin{itemize}
\item[${\dag}$] Work  supported   by  
U.S. Department of Energy grant DE-FG02-04ER41318, 
the Natural Sciences and Engineering research
council of Canada, the Fonds Nature et Technologies of Quebec.
G.T. thanks the Tomlinson foundation  for support.
S.J.~thanks RIKEN BNL Center
and U.S. Department of Energy [DE-AC02-98CH10886] for
providing facilities essential for the completion of this work.  
The authors wish to express their gratitude to Lucy Carruthers for invaluable assistance
in debugging and adapting the software described in this work.
\item[${\ddag}$] LPTHE, Univ.\,Paris 6 et 7
is: Unit\'e mixte de Recherche du CNRS, UMR7589.
\end{itemize}
}

\end{titlepage}

\tableofcontents

\vfill
\section*{Program Summary}
\vspace{0.3cm}

\noindent{\sl Title of the program:} \- {\tt SHAREv2}, \hfill April 2006
\vspace{0.3cm}

\noindent{\sl Computer:}\\
PC, Pentium III, 512MB RAM \hfill \lowercase{NOT HARDWARE DEPENDENT};
\vspace{0.3cm}

\noindent{\sl Operating system:}\\
Linux: RedHat 6.1, 7.2, FEDORA {\it etc.}  \hfill \lowercase{NOT  SYSTEM DEPENDENT};
\vspace{0.3cm}

\noindent{\sl Programming language:}\-
{\tt FORTRAN77}
\vspace{0.3cm}

\noindent{\sl Size of the package:}\- 
167 KB  directory, without libraries (see \\
http://wwwasdoc.web.cern.ch/wwwasdoc/minuit/minmain.html\\
http://wwwasd.web.cern.ch/wwwasd/cernlib.html\\
for details on library requirements  ).
\vspace{0.3cm}

\noindent{\sl Distribution format:}\- 
tar gzip file
\vspace{0.3cm}

\noindent{\sl Number of lines in distributed program (.tar file), including test data, {\it etc.}:}\- 
26017
\vspace{0.3cm}

\noindent{\sl Keywords:}\-
fluctuations, relativistic heavy-ion collisions, particle production, statistical models, 
decays of resonances
\vspace{0.3cm}

\noindent{\sl Computer:}\- 
Any computer with an f77 compiler
\vspace{0.3cm}

\noindent{\sl Nature of the physical problem:}\\
Event-by-event fluctuations have been recognized  to be the physical 
observable capable to  constrain particle production models.
Therefore, consideration of event-by-event fluctuations is required for 
a decisive falsification or constraining of (variants of) particle production models 
based on (grand-, micro-) canonical statistical mechanics phase space, the so 
called statistical hadronization  models (SHM). 

As in the case of particle yields, to properly compare model calculations to data 
it is necessary to consistently take into account resonance decays.
However,  event-by-event fluctuations are more sensitive than particle yields 
to experimental acceptance issues, and a range of
techniques needs to be implemented to extract `physical' 
fluctuations from an experimental event-by-event measurement.    
\vspace{0.3cm}

\noindent{\sl Method of solving the problem:}\\
The techniques used within the SHARE suite of programs \cite{share_orig} are
updated and extended to fluctuations.
A full particle data-table, decay tree, and set of experimental feed-down coefficients
are provided.    Unlike SHAREv1.x, 
experimental acceptance feed-down coefficients can be entered for {\em any} resonance decay.

SHAREv2 can calculate yields, fluctuations, and bulk properties of the 
fireball from provided thermal parameters;
alternatively, parameters can be obtained from fits to experimental data, 
via the MINUIT fitting algorithm \cite{minuit}.    
Fits can also be analyzed for significance,
parameter and data point sensitivity.

Averages and fluctuations at freeze-out of both the stable particles and the hadronic 
resonances are set according to a statistical prescription, 
calculated via a series of Bessel functions, using CERN library programs. 
We also have the option of including  finite particle widths of the 
resonances.    A $\chi^2$ minimization algorithm, also from the 
CERN library programs, is used to perform and analyze
the fit.   Please see \cite{share_orig} for more details on these.

\vspace{0.3cm}
\noindent{\sl Purpose:}

The vast amount of high quality soft hadron production data, 
from experiments running at the SPS, RHIC, in past at the AGS, 
and in the near future at the LHC,
offers the opportunity for statistical particle production 
model {\em falsification}. 
This task has turned out to be difficult when 
considering solely particle yields   addressed in 
the context of SHAREv1.x~\cite{share_orig}. For this reason  
{\em physical conditions at freeze-out} remain contested.
Inclusion in the analysis of event-by-event fluctuations appears
to resolve this issue.     Similarly, a thorough analysis including both 
fluctuations and average multiplicities gives a way to explore the
 presence and strength of  interactions following
hadronization (when hadrons form), 
ending with thermal freeze-out (when all interactions cease).

SHAREv2 with fluctuations will also help determine
which statistical ensemble (if any), e.g., canonical or grand-canonical, 
is more {\em physically} appropriate for analyzing a given system.
Together with resonances, fluctuations can also be used for a direct
estimate of the extent the system re-interacts between chemical and thermal 
freeze-out.

We hope and expect that 
SHAREv2 will contribute to decide if any of the statistical hadronization
model variants
has a genuine physical connection to hadron particle production.

\vspace{0.3cm}
\noindent{\sl Computation time survey:}

We encounter, in the FORTRAN version computation, times  up to  
 seconds for evaluation  of particle yields. These rise 
by up to a factor of  300 in the process of minimization
and a further factor of a few  when   $\chi^2/{\rm N_{DoF}}$ profiles and
 contours 
with  chemical non-equilibrium are requested. 
\vspace{0.3cm}

\noindent{\sl Accessibility: } 

The program is available from:
\begin{itemize}
\item The CPC program library,
\item The following website:\\
 {\tt http://www.physics.arizona.edu/$\tilde{\phantom{~}}$torrieri/SHARE/share.html }
\item from the authors upon request.
\end{itemize}

\noindent{\bf SUMMARY OF NEW FEATURES (w.r.t. SHAREv1.x)}
\vspace{0.2cm}

\noindent{\sl Fluctuations:} In addition to particle yields, ratios and bulk quantities
SHAREv2 can calculate, fit and analyze statistical fluctuations 
of particles and particle ratios;  

\vspace{0.2cm}

\noindent{\sl Decays:}  SHAREv2 has the flexibility to account for any
experimental method  of allowing for decay feed-downs to 
the particle  yields;
\vspace{0.2cm}

\noindent{\sl Charm flavor:}
Charmed particles have been added to the decay tree, allowing as an option study
of statistical hadronization of $J/\psi$,  $\chi_c$, $D_c$, etc.; 

\vspace{0.2cm}

\noindent{\sl Quark chemistry:}
Chemical non-equilibrium yields for both $u$ and $d$ flavors, as opposed to generically light quarks $q$, 
are considered;  $\eta$--$\eta'$  mixing, etc., are properly dealt with, and chemical non-equilibrium can 
be studied for each flavor separately;

\vspace{0.2cm}

\noindent{\sl Misc:}
Many new commands and features have been introduced and added to the basic user interface.
For example, it is possible to study combinations of particles and their ratios. 
It is also possible to combine all the input files into one file.

\vspace{0.3cm}
\noindent{\bf SHARE Compatibility and Manual:}\\
This write-up   is  an update and extension of \cite{share_orig}.  
The user should consult Ref.\cite{share_orig} regarding 
the principles of user interface and for all particle yield
related physics and program instructions, other than the parameter 
additions and  minor changes  described here.  SHAREv2 is downward 
compatible  for the changes of the user interface, offering the user 
of SHAREv1 a computer generated revised input files compatible 
with SHAREv2.
 
\newpage
\section{\label{intro} Introduction}

The statistical hadronization model \cite{Fer50,Pom51,Lan53,Hag65} (SHM) assumes particles are 
created according to their phase space weight, given the
locally available energy and quantum numbers. Such a reaction 
model implies that the underlying dynamics
of strong interactions saturates the strength of each particle production quantum matrix element.
 
This approach can be used to calculate the event-by-event average, as well as fluctuation 
(distribution width) and higher cumulants
of any `soft' observable.
Event-by-event particle fluctuations have been subject to  intense current
theoretical~\cite{fluct1,fluct2,fluct3,fluct4,fluct4b,fluct5,fluct6,fluct7,fluct8},
and experimental  interest~\cite{starfluct,starfluct2,phefluct,spsfluct}. SHAREv2 will offer a
standardized framework to evaluate these. 

While {\em qualitative} study of fluctuations is useful as a test of new physics, a quantitative analysis including both average multiplicities
and their event-by-event fluctuations constitutes a powerful probe of hadronization conditions 
\cite{mefluct1,mefluct2,mefluct3,mefluct4}.
In particular, the following questions can be addressed when both yields and fluctuations 
are considered in the same model framework: 
\begin{itemize}
\item SHM can be falsified if and when fluctuations do not scale w.r.t. averages as expected
in statistical physics. Moreover, only if the same set of thermal parameters gives
    good description of experimentally measured yields {\em and}
    fluctuations, can we claim that the SHM fit is physically sound. 
\item  As has recently been shown \cite{nogc1,nogc2,nogc3,becattini,steinberg}, the value
 to which the scaled variance $\sigma_N$ (see Eq.\,(\ref{fluctdef}))  for a single particle 
converges in the thermodynamic limit varies 
  by as much as an order of magnitude when different statistical
ensembles are considered. Thus, fluctuations can help decide if and when certain particle yields should 
be studied in grand canonical or canonical ensembles. 
\item SHM fits containing both the average particle multiplicity and the fluctuation  de-correlate the hadronization temperature $T$ and  light quark phase space occupancy $\gamma_q$ 
(see Eq. \ref{upsilons} and \cite{share_orig}) typical of 
fits when only the average multiplicities are fitted~\cite{mefluct1}.

Therefore, the study of both fluctuations and yields can help to experimentally distinguish between
the chemical equilibrium  freeze-out model ($T\simeq 170$ MeV, $\gamma_q=1$ \cite{equil_energy}),
or the best fit with  chemical non-equilibrium  at typically lower $T$ \cite{gammaq_energy}. 
\item Considering the directly detectable resonance decays, fluctuations of particle 
yield ratios offer a way 
to quantitatively gauge the effect of hadronic re-interactions between formation and thermal
freeze-out \cite{mefluct2}.
\end{itemize} 
 
To investigate these questions, it is necessary, in evaluation of both particle yields and fluctuations to:
\begin{itemize}
\item Incorporate all particles resonance decay trees \cite{pdg}
in the program structure;
\item Obtain    particle yields and fluctuations  for a
given set of thermodynamic parameters;
\item i) Check  if the  
parameters obtained by fitting particle yields are consistent with observed fluctuations;\\
ii)  Once all corrections to fluctuations due to experimental setup are
understood, incorporate the fluctuations along with yields into the chemical freeze-out 
fitting procedure.
 \end{itemize}
SHAREv2   comprises a framework that addresses these challenges.
 
As implied above,  event-by-event particle yield fluctuations are subject to many subtle 
experimental effects   which need to be understood and 
kept under control for a joint yield-fluctuation  analysis to proceed.   Further,
there is the  choice of statistical model ensemble in computation of the
phase space volume:\\ 
1) Evaluation with exact energy and discrete quantum number 
conservation (micro-canonical ensemble --- MCE);\\
2) In the canonical ensemble (CE), statistical energy fluctuations are allowed,   conserving
 discrete quantum number(s) exactly;\\
3) In the  grand-canonical ensemble (GCE),  statistical fluctuations of all conserved 
quantities occur --- there are also mixed CE-GCE ensembles where some particle yields are 
conserved and other fluctuate.

 Clearly, the fluctuations of particle yields are most constrained in MCE
    and least constrained in GCE.
Thus, although in  the three ensembles, the first moments of 
any observable distribution, i.e., expectation values,  
coincide in the thermodynamic limit, this will
not be the case for the fluctuations\cite{nogc1,nogc2,nogc3}. The choice of appropriate ensemble 
in the situation considered has to be made based on evaluation of prevailing 
{\em physical} conditions. 

In study of total particle yields, in the physical context of heavy ion collisions,   
the electrical charge and baryon number are fixed and, in these variables, we have to 
consider the CE or MCE if and when we are observing all particles. On the other hand, 
if we  only observe a sub-volume of the  system, which is exchanging energy and particles 
with an unobserved `bath' consisting of the remainder of the reaction system, 
then, also   {\em  conserved} quantum numbers  
must be allowed to fluctuate, which implies use of the GCE for all observables.
However, when the totality of the produced particles carrying a conserved quantum 
number falls within the detector acceptance region, occasional detection of one such particle implies presence 
of the corresponding anti-particle, and thus in that case CE or MCE must be applied.

Within the context of heavy ion physics, with reactions occurring at large energy,
a study of fluctuations within a narrow momentum rapidity\footnote{\noindent 
The rapidity $y$, defined by $y=\frac{1}{2} \ln\left( \frac{E+p_z}{E-p_z} \right)$, 
where $E$ is the particle energy and $p_z$ the particle 
momentum component parallel to  the collision axis, is a variable additive under 
Lorentz transformations parallel to this axis.} acceptance 
window provides for the division
between `system' and `bath', with the bath being the  unobserved rapidity domain.  
In all experiments currently capable to measure fluctuations, 
detector acceptance is limited typically
to the central rapidity {\em phase space} coverage. Such an acceptance  domain
 in the boost invariant (denoted below as subscript b.i.) 
limit is  equivalent  to a {\em configuration space} sub-volume\cite{cleymans} 
and thus for both particle ratios,  and particle yield  width  
(fluctuation)\cite{mefluct4}, we have:
\begin{equation}
\frac{\ave{N_i}_{\rm GC}}{ \ave{N_j}_{\rm GC}}
=
\frac{(dN_i/dy)_{\rm b.i.} }{ (dN_j/dy)_{\rm b.i.}}.
\label{eq:boost_inv}
\end{equation}
Here, $\ave{N_i}$ is the event-by-event average of particle $i$, $dN_i/dy$ is the average number 
of particles in an element of rapidity at central rapidity.

Similarly, barring acceptance effects (discussed in sections \ref{secweakdyn} and \ref{secweakcor} of this work, 
as well as in \cite{mefluct4}), the scaled variance, defined as
\begin{equation}
\label{fluctdef}
\sigma_X^2 = \frac{\ave{(\Delta X)^2}}{\ave{X}} = \frac{\ave{X^2}-\ave{X}^2}{\ave{X}},
\end{equation}
will be given by
\begin{equation}
\sigma^2_{N_i}
=
\left( \frac{d \sigma^2_{N_i}}{dy} \right)_{\rm b.i.}.
\end{equation}
We conclude that, in experiments with limited central rapidity acceptance,   both yields and 
fluctuations  should be evaluated in the GCE with respect to the conserved 
quantum numbers (charge $Q$ , baryon number $b$,
strangeness $s-\bar s$).

Use of the GCE, for at least some conserved charges, is {\em required} 
by the experimental observation of a significant fluctuation {\em in those charges} \cite{starfluct,starfluct2,phefluct}.  
 This fluctuation has been found to be compatible with
Poisson scaling,
\begin{equation}
\label{poisson}
\ave{(\Delta N)^2} \sim \ave{N},
\end{equation}
which is approximately followed by the GCE fluctuations. This is not the only scaling
known to be present in this area of physics.   
Elementary reaction systems have been observed to follow a
non-Poissonian  scaling \cite{heiselberg,kno} w.r.t. multiplicity averages,
\begin{equation}
\label{kno}
\ave{(\Delta N)^2} \sim \ave{N} + c \ave{N}^2,
\end{equation}
where $c$ is a constant.
As has been argued previously, \cite{freje1,freje2,freje3}, it is possible to describe this scaling
by considering an extension of the Grand Canonical ensemble  
(variously referred to as Isobaric or Pressure ensemble) where system volume 
is also allowed to fluctuate.

In SHAREv2, we   consider only GCE yields and fluctuations and search to explore whether 
the  grand canonical statistical hadronization model can quantitatively reproduce fluctuations 
in the same way as it was shown to reproduce particle yields in heavy ion A--A reactions.

\subsection{Evaluation of yields and fluctuations}

In GCE,  particle yields and   fluctuations can be calculated by a textbook
method \cite{huang}.   For a hadron with an energy $E_{p} = \sqrt{p^2+m_i^2}$,  the energy 
state occupancy is, 
\begin{eqnarray}
n_{i}(E_p) = \frac{1}{  \Upsilon_i^{-1} e^{E_p/T}\pm 1},
\label{eqn}
\end{eqnarray}
where the upper sign is for fermions and the lower sign is for  bosons.   
The chemical fugacity $\Upsilon_i$ will be considered in section \ref{upsilon}.

The yield average  is obtained   by multiplying the occupancy number  Eq.\,\ref{eqn}
by the density of states (where $V$ is volume and $g$ degeneracy): 
\be
\label{yield_formula}
\langle N_i\rangle
 & = & gV\int \frac{d^3p}{ (2\pi)^3}\, n_{i}(E_p).
\ee
The  fluctuation in this number is found to be:
\be
\label{fluct_formula}
\ave{(\Delta N_i)^2}
=   \Upsilon_i \left.{\partial\ave{ N_i}\over \partial \Upsilon_i} \right|_{T,V}=
gV\int \frac{d^3p}{ (2\pi)^3}\, n_{i}(E_p) \left(1 \mp n_i(E_p)\right).
\end{eqnarray}
Eqs.\,(\ref{yield_formula}--\ref{fluct_formula}) can be
 evaluated to any desired accuracy by
converting them into an expansion of Bessel function terms
\cite{Lan53}, 
\begin{eqnarray}
\label{yield_formulaK}
\langle N_i\rangle  & = & \frac{ gV T^3}{2\pi^2}
                 \sum_{n=1}^{\infty} \frac{(\pm 1)^{n-1} \Upsilon_i^n}{n^3} 
                  W \left(\frac{n m_i}{T}  \right), \qquad \Upsilon_i e^{-m_i/T}<1,\\
\label{fluct_formulaK}
\ave{(\Delta N_i)^2}  & = & \frac{ gV T^3}{2\pi^2}\sum_{n=1}^{\infty}  \frac{(\pm 1)^{n-1} \Upsilon_i^n}{n^3}
        \left(\begin{array}{c}  2+n-1 \\   n \end{array}  \right)  W\left( \frac{n m_i}{T} \right), 
\end{eqnarray}
where $W(x)=x^2 K_2(x)$ (see \cite{share_orig}, section 2, for the technical details 
required in doing these calculations, as well as a discussion of particles 
with finite width).
\subsection{Chemical potential and chemical non-equilibrium \label{upsilon}}
The particle fugacity $\Upsilon_i$ can be obtained from the quark number
content of the particle as well as the fugacities and phase space occupancies of the individual quark flavors.
If a particle has $N_q=N^i_u+N^i_d$, $N^i_s$ and $N^i_c$, light up + down,
 strange and charm quarks,   
$N^i_{\overline{q}}=N^i_{\overline{u}}+N^i_{\overline{d}}$, $N^i_{\overline{s}}$ 
and $N^i_{\overline{c}}$ antiquarks,  and isospin $I_3$,
the fugacity $\Upsilon_i$, or, equivalently the associated particle chemical 
potential $\mu_i$ will be given by:
\begin{equation}
\Upsilon_i=e^{\mu_i/T}=
\left(\lambda_u \gamma_u\right)^{N^i_u}
\left(\lambda_d \gamma_d\right)^{N^i_d}
\left(\lambda_s \gamma_s\right)^{N^i_s}
\left(\lambda_{c} \gamma_{c}\right)^{N^i_{c}}
\left(\lambda_{\bar u} \gamma_{\bar u}\right)^{N^i_{\bar u}}
\left(\lambda_{\bar d} \gamma_{\bar d}\right)^{N^i_{\bar d}}
\left(\lambda_{\bar s} \gamma_{\bar s}\right)^{N^i_{\bar s}}
\left(\lambda_{\bar c} \gamma_{\bar c}\right)^{N^i_{\bar c}},
\label{upsilons}
\end{equation}
where $\lambda_{\overline{i}}= \lambda_{i}^{-1}$,  and $\gamma_{\overline{i}}=\gamma_i$.
The individual $u$, $d$ light quark variables are related to the SHARE $q$ and $I_3$
variables $\lambda_q=\sqrt{\lambda_u\lambda_d}$ and $\lambda_{I_3}=\lambda_u/\lambda_d$,
see Ref.~\cite{share_orig}, and similarly 
$\gamma_q=\sqrt{\gamma_u\gamma_d}$ and $\gamma_{3}=\gamma_u/\gamma_d$,
see Eq.(\ref{gam3eq}) below.
The condition of chemical equilibrium, for a flavor $f$, imposes
$\gamma_f=1$ \cite{huang}.  The assumption of chemical equilibrium is not automatic 
in a dynamically expanding system
with a possible phase transition and, in fact, good theoretical arguments have been 
proposed for $\gamma_{q,s} \ne 1$
for a range of energies (see \cite{gammaq_energy}, and references therein).
However, it is difficult, using fits to particle yields, 
to distinguish between two models based on different
temperatures and $\gamma$ values. 

For instance, models based on both a higher freeze-out temperature and $\gamma_q=1$ \cite{equil_energy} 
or a lower freeze-out temperature and $\gamma_q>1$ \cite{gammaq_energy} have been used to fit 
SPS and RHIC data.   As shown in \cite{mefluct4}, this ambiguity is resolved when both yields and
fluctuations can be  considered.                                          

A complication arises for hadrons such as the $\pi^0$ or the $\eta$, which are in a flavor superposition
state.   If $\gamma_{u,d,s} \ne 1$, the yield of the 
hadron with fractional flavor content is considerably altered by the mixing.
For a meson of fractional quark number structure,
 \begin{equation}
|i>=\alpha_u |u \overline{u}>+\alpha_d |d \overline{d}>+\alpha_s |s \overline{s}>, \phantom{....}
\alpha_u^2+\alpha_d^2+\alpha_s^2=1,
\end{equation}
the fugacity comprises the chemical yield fugacities as follows:
\begin{equation}
\Upsilon_i = \lambda_i \left(\gamma_u^2 \alpha_u^2+\gamma_d^2 \alpha_d^2+\gamma_s^2 \alpha_s^2  \right).
\end{equation}
Fractional flavor content   has non-negligible influence 
on the abundances of $\eta^0$ and $\eta'$, and their decay products, in fits which 
allow for chemical non-equilibrium factor $\gamma_s$. The same remarks applies when
 $\gamma_3 = \gamma_u \gamma_d^{-1} \ne 1$ to  $\pi^0,\ \rho^0$, etc. Thus, $\gamma_3 \ne 1$ can 
considerably enhance $\pi^0\propto \gamma_3^2 + \gamma_3^{-2}$ yield, while 
maintaining $\pi^\pm$ yields symmetry (since $\gamma_3$ cancels out in $\pi^{\pm}$).

Importantly, the evolution of quark-coalesced hadrons into final quark-eigenstates 
hadrons (like the oscillation of neutral kaons into K$_S$ and K$_L$) 
 means that the `source' QGP quark content will {\em not}, in general, 
be equal to the `final' hadron quark content.  
\subsection{Resonance decays \label{resdecays}}
Eqs.\,\ref{yield_formula} and \ref{fluct_formula} can be used to calculate the event-by-event averages 
and fluctuations of all hadrons \textit{at hadronization}.
This, however, is quite different from the \textit{observed} averages and fluctuations, 
since most hadrons are strong resonances (unstable states), which decay
after freeze-out, either to stable particles or to other resonances.
The final state  particle yields can be computed by taking the effect of these feed-downs 
into account \cite{Hag65}.

The ensemble average of the total yield $\ave{N_i}$ is:
\be
 \label{resoyield}
\langle N_i\rangle_{\rm total} & = &
\langle N_i\rangle_{\rm direct} + \sum_{j\ne i}
B_{j \rightarrow i}  \langle   N_j \rangle .
 \label{fluctres}
 \ee
$B_{j \rightarrow i}$ is the probability (branching ratio)
for the decay products of $j$ to include~$i$.

The fluctuation after resonance feed-down is given by:
\begin{equation}
\ave{(\Delta N_{j\to i})^2}
 = 
 B_{j \rightarrow i}(\mathcal{N}_{j \rightarrow i}-B_{j \rightarrow i})\ave{N_j}
+
B_{j \rightarrow i}^2 \ave{(\Delta N_j)^2}.
\label{resofluct}
\end{equation}
The second term corresponds to the fluctuation in \textit{the yield of resonances}.
The first term, in the \textit{number of} $j \rightarrow i$ decays given 
the branching ratio $b_{j \rightarrow i}$.
$\mathcal{N}_{j \rightarrow i}$ is the number of particles type $i$ 
produced in the decay, so that 
$\sum_i B_{j \rightarrow i}=\mathcal{N}_{j \rightarrow i}$. \\
$\mathcal{N}_{j \rightarrow i}=1$ for nearly all decays of nearly all resonances;  The most significant
exception are decays to multiple $\pi^0$ s, such as $\eta \rightarrow 3 \pi^0$. 

\subsection{Volume fluctuations and fluctuations of ratios \label{fluctratios}}
The expression \ref{fluct_formulaK} neglects volume fluctuations, coming from centrality cuts and
dynamics of system expansion.    These are accounted for by dividing the observed 
fluctuation into an extensive and an intensive part,
\be \label{volfluctdef}
\ave{(\Delta X)^2} \approx
\ave{(\Delta x)^2}\ave{V}^2 + \ave{x}^2 \ave{(\Delta V)^2},
\ee
 $\langle x\rangle$, $\langle x^2\rangle$ can be calculated by the 
statistical methods described in this section.

It is difficult to describe the volume fluctuation coefficient $\ave{(\Delta V)^2}$  
in a model-independent way.
The most straight forward way to deal with this problem is to choose 
observables insensitive to $\ave{(\Delta V)^2}$.

Any observable where $ \ave{x}^2 \ll \ave{\Delta x)^2}$ would be 
a good candidate.  This is why the
 fluctuation  in electromagnetic charge has long been considered 
to be a promising observable \cite{fluct4}.

A more general approach is to consider the event-by-event fluctuation of 
particle ratios  \cite{fluct3}, where the volume fluctuation 
$\ave{(\Delta V)^2}$ is zero by construction.
 Fluctuation of particles ratios can be calculated from the denominator's and 
numerator's fluctuation once the full resonance decay tree is known \cite{fluct3}.
Note that, unlike in the case of particle yields, resonance decays produce both 
fluctuations and \textit{correlations}, since
a resonance can decay both into a numerator and a denominator particle.  
If this is the case, a high resonance admixture can
considerably reduce the fluctuation of a ratio w.r.t. Poisson expectation.

The formulas to be used are, for the event-by-event fluctuation of  the ratio of two particles $N_1/N_2$ (for example $K^-/\pi^+$ \cite{fluct3}):
\begin{equation}
\sigma^2_{N_1/N_2} = \frac{\ave{(\Delta N_1)^2}}{\ave{N_1}^2} 
+ \frac{\ave{(\Delta N_2)^2}}{\ave{N_2}^2} 
- 2 \frac{\ave{\Delta N_1 \Delta N_2}}{\ave{N_1}\ave{ N_2}}.
\label{ratiofluct}
\end{equation}
The last correlation term is given by the resonance decay into both particles
\begin{equation}
\ave{\Delta N_1 \Delta N_2}  =  \ave{N_1 N_2} - \ave{N_1}\ave{N_2} \simeq
\sum_{j } B_{j \rightarrow 1,2} \ave{N}_j.
\end{equation}
$\sigma^2_N$ does not depend on the {\em average} system volume  ($\ave{V}$), 
since it cancels between the numerator and the denominator.
$\sigma^2_{N1/N2}$, however, 
does acquire a dependence  on $\ave{V}$ since ratio fluctuations scale as $\ave{N}^{-1}$.
Hence, an analysis incorporating fluctuations of particle ratios should also consistently account
 for particle yields, and the system normalization
(thermodynamic parameter {\bf norm}, \cite{share_orig} section 3.1) should be considered 
as a fit parameter.

The equations presented in this section can be used to compute fluctuations of 
particles yields and ratios from given SHM parameters (temperatures, chemical
potentials, and phase space occupancies).
However, it has long been known \cite{pruneau} that fluctuations are 
considerably less robust than yields against systematic effects resulting 
from limited experimental acceptance.   These effects, therefore, 
have to be taken into account \textit{within} the SHM.  
The next two subsections give two such issues addressed within SHAREv2.

\subsection{Fluctuations and detector acceptance \label{secweakdyn}}
One way to separate detector acceptance effects from 
physics is to eliminate the former via mixed event techniques.   
A `static' fluctuation $\sigma^2_{stat}$ is measured 
in a sample of fake events, constructed by
using tracks from different events \cite{starfluct}.   
Since tracks from different events have no correlations 
or quantum corrections, $\sigma^2_{stat}$ is determined 
solely by a trivial Poisson contribution as well 
as detector acceptance effects.

Within the statistical hadronization model:
\begin{equation}
\label{poisson_single}
(\sigma^2_{N_i})_{stat}= 1\,.
\end{equation}
For particle ratios in mixed events, the correlation 
term, $\ave{\Delta N_1 \Delta N_2}$ in Eq.\ref{ratiofluct}, vanishes,
while $\ave{(\Delta N_{1,2})^2}$ follow Poisson scaling.
Hence, Eq.\,\ref{ratiofluct} reduces to
\begin{equation}
(\sigma^2_{N_i/N_j})_{stat} = \frac{1}{\ave{N_i}}+\frac{1}{\ave{N_j}}.
\end{equation}
The `dynamical fluctuation' $\sigma^2_{dyn}$ 
\cite{pruneau,Gazdzicki:1992ri,supriya,fluct3} 
corresponds to the difference between the `raw' 
total fluctuation $\sigma^2$ and the fake event fluctuation:   
\begin{equation}
\sigma^2_{dyn} \equiv \sqrt{\sigma^2-\sigma^2_{stat}}. \\
\end{equation}
In certain limits, $\sigma^2_{dyn}$
can be shown \cite{pruneau} to be independent of detector acceptance.   
Hence, comparing SHM estimates of $\sigma_{dyn}$ to experimental 
measurements is more reliable than using $\sigma^2$.

\subsection{Correlations, resonances and detector acceptance \label{secweakcor} }
Because mixed event tracks are uncorrelated, mixed event techniques cannot account 
for detector acceptance effects within {\em particle correlations}.
Thus, the branching ratios appearing in Eqs.\,\ref{resoyield} and \ref{resofluct}
need to be modified,  $B_{j \rightarrow i} \rightarrow \alpha_{j \rightarrow i} B_{j \rightarrow i} $, 
and in addition
Eq.\,\ref{ratiofluct} needs to be updated,
\begin{equation}
\sigma^2_{N_1/N_2} = \frac{\ave{(\Delta N_1)^2}}{\ave{N_1}^2} 
+ \frac{\ave{(\Delta N_2)^2}}{\ave{N_2}^2} - 
2 \alpha_{12} \frac{\ave{\Delta N_1 \Delta N_2}}{\ave{N_1}\ave{ N_2}},
\label{ratiofluct_acc}
\end{equation}
here $\alpha_{j \rightarrow i}$ refers to the probability that particle $i$ will end up
in the detector's acceptance region given particle $j$ is in the region, while $\alpha_{12}$ measures the probability that {\em both} decay
products will be inside this region.

For a boost invariant azimuthally complete system, $\alpha_{j \rightarrow i}=1$ 
since particles leaving the detector
acceptance region will be balanced by particles entering.   
However, in general $\alpha_{12}<1$, since
if a resonance is outside the detector acceptance region 
{\em both} particles can {\em not be} inside it, and
the intrinsic particle decay momentum adds a rapidity scale 
to the system, breaking boost invariance \cite{mefluct1}.

See \cite{mefluct1} for an illustration of how to calculate 
$\alpha_{12}$.   While such a comprehensive calculation 
is outside the scope of the 
current version of the program,  we offer the user the possibility of 
entering an $\alpha_{12}$ for any resonance decay as an input parameter. 
See section \ref{weakdecaysect} on how to do this.
In practice, this should only be necessary for a few most frequent
and energetic resonance decays, such as 
$\rho \rightarrow \pi \pi$ and $K^* \rightarrow K \pi$.

\section{Implementation of GCE fluctuations in SHAREv2 \label{fluctshare}}
Experimental event-by-event fluctuation data points were implemented in the SHARE interface 
in a similar manner as yield and ratio data points (see \cite{share_orig}, section 3.4).
The tag which denotes that a fluctuation is being calculated is {\bf fluct$\_$yld}.
A statement such as

\begin{tabular}{llllll}
\noindent {\bf particle1}& {\bf fluct$\_$yld}  & {\bf  data} & {\bf $\Delta_{\rm stat}$}&   
                           {\bf $\Delta_{\rm syst}$} &   {\bf fit?}
\end{tabular}\\
will calculate $\sigma^2_N$ of {\bf particle1} (defined in Eq.\,\ref{fluctdef}).
 
If {\bf fit?} is set to 1, this data point is used within a fit together with the
experimentally measured value {\bf data} and the
statistical  ($\Delta_{\rm stat}$) and systematic ($\Delta_{\rm syst}$) error .   
The format of the data-line is exactly the same as in SHAREv1.x
(\cite{share_orig} section 3.4).

To calculate the fluctuation of a ratio, {\bf  particle1} should 
be substituted by the data point number where the ratio is defined.
For instance, if the 5th data point (from the top) is a $K^-/\pi^-$ 
ratio, than it's fluctuation is given by:

\begin{tabular}{llllll}
\noindent {\bf 05$\phantom{XXXX}$    }& {\bf fluct$\_$yld}  & {\bf  data} & 
  {\bf $\Delta_{\rm stat}$}&   {\bf $\Delta_{\rm syst}$} &  {\bf fit?}\\
\end{tabular}\\
(See section \ref{combo} of this paper for a more general treatment of this data point referencing).

SHAREv2 implements most definitions of dynamical fluctuations used to date by experimental collaborations.
These are implemented as additional tags of {\bf fluct$\_$xxx} type, where {\bf xxx} 
refers to different ways the experimental
measurement is presented.

The possible types of data points are:
\begin{description}
\item[fluct$\_$dyn] To calculate $\sigma_{dyn}=\sqrt{\sigma^2-\sigma^{2}_{stat}}$, as measured in \cite{supriya},
\item[fluct$\_$dnr] To calculate $\sigma_{dyn}=\displaystyle\frac{\sigma^2}{\sigma^{2}_{stat}}$ as suggested in \cite{fluct3}. 
\end{description}

\section{Decay feed-down and particle yields \label{weakdecaysect}}
\subsection{Particle decay acceptance data files}
As shown in section \ref{intro}, decay feed-down is a fundamental component of the statistical hadronization model.  
However, the limited coverage of most detectors means that the feed-down coefficients 
will acquire an \textit{experimental}
correction, corresponding to the probability that the decay products of a given resonance formed 
within the detector acceptance region will also be in that 
region.   As shown in section \ref{secweakcor}, these corrections need to be considered when 
calculating both fluctuations and yields.

Weak decays, such as $\Lambda \rightarrow p \pi^-$ (most protons at RHIC are, in fact, given by feed-down from hyperons), 
are particularly susceptible to experimental acceptance, 
as they occur at a \textit{macroscopic} distance from the primary vertex.   
Hence, weak experimental feed-down corrections include a geometrical as well as a momentum space component.     

Since the `parent' particles are not always directly observed, SHARE must 
be able to compute final hadron multiplicities including experimental feed-down coefficients 
for all decays where this effect is non-negligible.

SHAREv1.x allowed the user to input experimental (weak) feed-down contributions
to produced particle yields via four acceptance coefficients:\\
{$K_S \rightarrow {\rm \, anything}$, $K_L \rightarrow {\rm \, anything}$,
$Y\rightarrow$ Mesons, $Y \rightarrow$ baryons} 
(see \cite{share_orig}, section 3.4.1).

It turns out this approach was not sufficiently flexible: 
for instance, undetected $\Sigma \rightarrow N \pi$ feed-down can be   
treated very differently from $\Lambda \rightarrow p \pi^-$ corrections,
considering the difference in lifespan,  and (vertex)  acceptance cuts applied.
 
Moreover, the experimental acceptance of
different {hyperon $\rightarrow$ nucleon} weak decays, such as 
 $\Xi^- \rightarrow \Lambda \pi^-$ as compared to   $\Lambda \rightarrow p \pi^-$ 
is likely to be considerably different. Finally, 
different weak decays of the same hadron can have varying acceptances, compare
  $K_L \rightarrow 3 \pi$, with  $K_L \rightarrow \pi e \nu$, 
and with $K_L \rightarrow \pi \mu \nu$.
A similar acceptance problem may arise in special cases
involving   strong decay chains when the acceptance region is particularly narrow and/or the particles rapidity distribution is not well understood. 

A more flexible way of treating weak 
decay contributions to particle yields is therefore necessary.
Specifically, there should be an easy way to  
allow for any {\em arbitrary} decay/reaction contributing to {\em any} data point.
SHAREv2 provides such a possibility through user defined decay feed-down files.

In data file containing the experimental results  to be fitted
 (see \cite{share_orig} section 3.4.3), a weak 
decay control file  is now signaled by a statement of the type:\\
{\bf Weakdecay   File.feed}\\
where {\bf File.feed} is a 9-letter filename.
The program   then obtains the   decay acceptance weights from {\bf File.feed},
an ASCII file in a format similar to the decay 
tree files (described in \cite{share_orig} section~3). 

\begin{figure}[t]
\centering
  \epsfig{width=17cm,figure=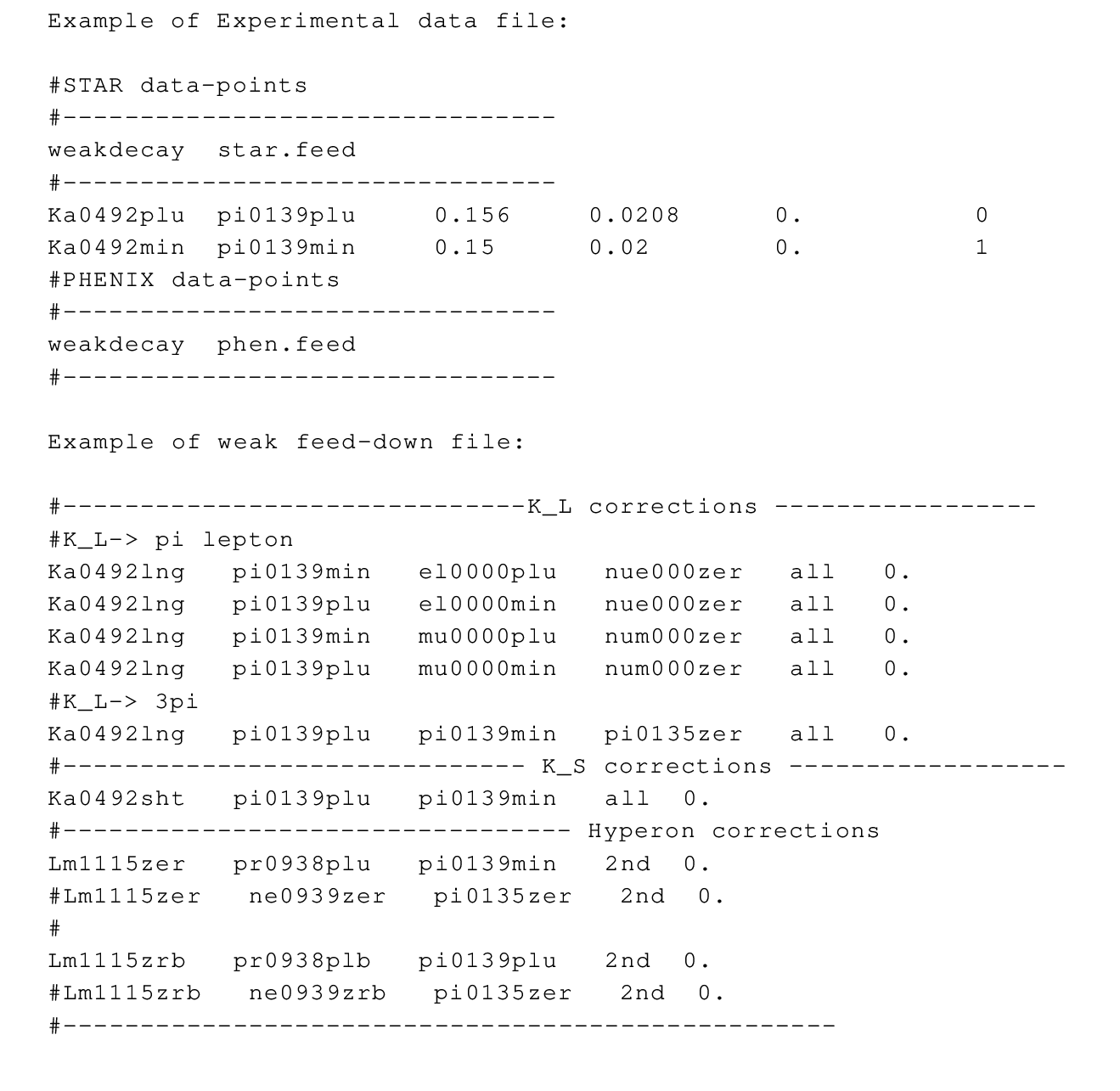}
 \caption{\label{weakfeedfig} An example of the SHAREv2 weak feed-down acceptance coefficient
implementation.}
\end{figure}

 Fig. \ref{weakfeedfig}, and the attached input files provided with the 
SHARE package, show how to implement the weak decay acceptance coefficients.  
While many weak feed-down files might be involved 
in the same analysis, generally, they are experiment-specific, 
and hence can be kept track of in a systematic way.  
Alternatively, all weak feed-down files and experimental
data-files can be combined in a single large file, using 
the methods described in section \ref{singlefile} of this paper.

In more detail, 
a typical line in a feed-down file will be:

\begin{tabular}{llllll}
{\bf Parent}&  {\bf Daughter$_1$}& {\bf Daughter$_2$} & 
                  {\bf all/1st/2nd/cor} & {\bf coeff}\\
\end{tabular}

\noindent or, for 3-body decays, 

\begin{tabular}{llllll}
{\bf Parent }& {\bf Daughter$_1$}& {\bf Daughter$_2$} & {\bf Daughter$_3$} &
               {\bf  all/1st/2nd/3rd/cor}& {\bf coeff}\\
\end{tabular}

\noindent The switch {\bf all/1st/2nd/3rd} refers to the daughter 
to which the decay coefficient applies.

{\bf all} means that the decay coefficient is the same for all 
daughters, while {\bf 1st/2nd/3rd} means only the 1st/2nd/3rd daughter will be removed 
from the experimental yield.
For example, in the $\Lambda \rightarrow \pi p$ decay in STAR \cite{barannikova,starlambda},  
STAR accepts the nucleon from the $\Lambda$ decay but not the $\pi$, and this fine tuning of
the decay is clearly quite important as a relatively large fraction of all nucleons comes from 
weak $\Lambda$ decays.

{\bf cor} refers to the fractional contribution of the acceptance to the two particle correlation\\
$\langle Daughter_1$ $ Daughter_2\rangle $ 
induced by a common resonance decay from parent, 
denoted as $\alpha_{12}$ in Eq.\,\ref{ratiofluct_acc} (section \ref{secweakcor}).

SHARE will renormalize the decay
{\bf Parent } $\rightarrow$ {\bf  all/1st/2nd/3rd/cor}\\
by the coefficient {\bf coeff} when calculating all data points after the given {\bf weakdecay} statement. 

It is therefore possible to assign a different weak decay file to each data point, 
or assume that a group of data points are subject 
to the same set of weak decay yield contribution,
(e.g., in general all data points from the same experiment should have the same weak decay file).
The way to do this is the same as in v1.x (\cite{share_orig} section 3.4.3):  
when the program reads a {\bf weakdecay} statement, it assigns the current decay pattern
to {\em each} data point encountered until  a {\em new} {\bf weakdecay} statement is met.

Two special case exist, for which no {\bf File.feed} file is needed:\\
{\bf weakdecay   UNCORRECT} `uncorrected' (from the perspective of experimental data set) 
means that   all weak decays contributions to particle yields are  fully accepted by  SHAREv2. \\  
{\bf weakdecay   NOWK$\_$FEED} means that {\em all} particle yields are computed
{\em without} contributions from weak decays, from the perspective of experimental data 
this means that either all weak decay products are {\em not} accepted and/or have been 
all corrected for in experimental yields, as, e.g., applies to some NA49 results.

When fluctuations are considered, it is important to deal carefully with experimental 
corrections, which are neither close to total (close 100 \% accepted) nor null
(close to 0 \% accepted).  Weak decay corrections of the daughter 
particles are usually correlated with each other in momentum space, 
so the straight-forward application of 
Eq.\,\ref{fluct_formula} will not be a good description of 
fluctuations with a non-trivial detector acceptance function \cite{fluct6}.
In this case, it is better to use dynamical, rather than total, 
fluctuations as discussed in section \ref{secweakdyn}.
 
\subsection{Compatibility with SHAREv1.x experimental data files}
The improved weak decay treatment does not impair compatibility 
of experimental input files between SHAREv2 and SHAREv1.x.
SHAREv2 will read a SHAREv1.x experimental data-file, and 
automatically calculate  applicable contributions for each weak decay based on the
information contained in the SHAREv1.x {\bf weakdecay} statement.   
A line will be printed within the {\bf sharerun.out} output file that
signals a SHAREv1.x format {\bf weakdecay} statement was encountered.

In addition, an output v2 weakdecay file called {\bf weak\#v1.x} 
(where {\bf \#} refers to the data point number) is automatically 
generated translating the v1.x weak decay information into v2 format.
The user is advised to eventually change all {\bf weakdecay} lines to\\
{\bf weakdecay  weak\#v1.x}\\
 as the v2 format is considerably more powerful and less
amenable to systematic error stemming from an incomplete 
understanding of weak decays.

\section{\label{chemsec}Quark chemistry}
SHAREv1.x input files listed particle chemical content by total isospin $I$ and its third component, $I_3$,
as well as the number of light ($q$, either $u$ or $d$), strange ($s$) and charm ($c$) quarks
(See section 3.2 of \cite{share_orig}).
In SHAREv2, $u$ and $d$ quarks are now separately accounted for. The particle listing format is: 

\noindent {\bf name\quad mass\quad  width\quad  spin\quad  I\quad  
I3\quad  u\quad d\quad s\quad  au\quad ad\quad as\quad  c\quad   ac\quad  MC}

\noindent where {\bf name} is the particle's 9-character name, {\bf $I$} 
and {\bf $I_3$} are the total and third component of the isospin,
{\bf u, d, s, c} are the numbers of up, down, strange and charm quark numbers, 
while {\bf au, ad, as, ac} are the respective anti quark numbers.
The format of the table is otherwise identical to that discussed in \cite{share_orig}, section 3.2.

To check for the possibility 
that phase space occupancy differs  for the up and down quarks, 
a statistical model fit parameter (see \cite{share_orig}, section 3.1)
{\bf gam3} ($\gamma_3$) has been introduced, such that:
\begin{equation}\label{gam3eq}
\gamma_u = \gamma_q\gamma_3, \phantom{......................}\gamma_d = \gamma_q/\gamma_3.
\end{equation}
The quark/anti quark numbers can be fractional, to account for the superposition states 
described at the end of section \ref{upsilon}.

To calculate ($u,\ d,\ s$) 
quark abundance in the  
statistically hadronizing QGP system (different, in general, from the freeze-out content, 
as shown in section \ref{upsilon}),
 new bulk variables {\bf tot$\_$u$\_$qgp}, 
{\bf tot$\_$d$\_$qgp} and {\bf tot$\_$s$\_$qgp} were introduced.  
These can be used in the same way as other bulk variables (see
\cite{share_orig}, section 3.4.2).
\subsection{Charm mesons}
Charmed particles have now been added in the files {\bf particles.data}
and {\bf partnowdt.data}.
Their nomenclature follows the general structure as described in  \cite{share_orig}, section 3.2.
   {\bf Dcxxxxxxx} refer to $D_c$ mesons, {\bf Dsxxxxxxx}, 
{\bf chixxxxcc} to $\chi_c$ states and {\bf psixxxxcc} to $J/\psi$ states.

Their abundance is regulated by the chemical potential $\lambda_c$ and the phase space occupancy $\gamma_c$, 
described in \cite{share_orig}, section 3.1.

\section{User Interface files and new commands \label{miscsec} }
\subsection{New single  file control \label{singlefile}}
As described in detail in \cite{share_orig}, section 3, SHARE relies on quite a few input files.
\begin{itemize}
\item The run-file {\bf sharerun.data}
\item The particles list
\item The particles decay tree
\item The initial values of the thermodynamic parameters
\item The experimental data points
\item Initialization for each fit parameter
\end{itemize}
This structure makes it easy to quickly explore regions of parameter space 
within an analysis {\em in progress}.  However, this
system makes it easy to mistakenly lose
a successfully completed and saved analysis,  since a change
in each of the files could considerably alter the end result.  
The introduction of weak decay correction file (see section
\ref{weakdecaysect}) aggravates
this problem.

SHAREv2, therefore, makes it possible to combine some, or all, input files into a single file.
Once the user found an optimum analysis, all input files involved in it can be 
combined into one large {\bf sharerun.data} file,
which can be easily kept
for future reproduction and modification.

This is done by changing the extension ({\bf.data} or {\bf.feed}) of the 
filename into {\bf.HERE}.  If the program encounters a filename ending 
in {\bf .HERE}, it assumes the relevant input is immediately
following the given line within the currently read file.   
The subsequent format is assumed to be {\em unchanged} from what 
it would have been had a separate file been opened (comments, etc.).
The only difference is that a {\bf *} symbol on a new line has to be 
present at the point where the separate file would have {\em ended}. 
When the program encounters the {\bf *} symbol, it switches back to reading the `earlier' file,
that is prior to the insert {\bf .HERE}.

\begin{figure}[t]
\centering
  \epsfig{width=16cm,clip=,figure=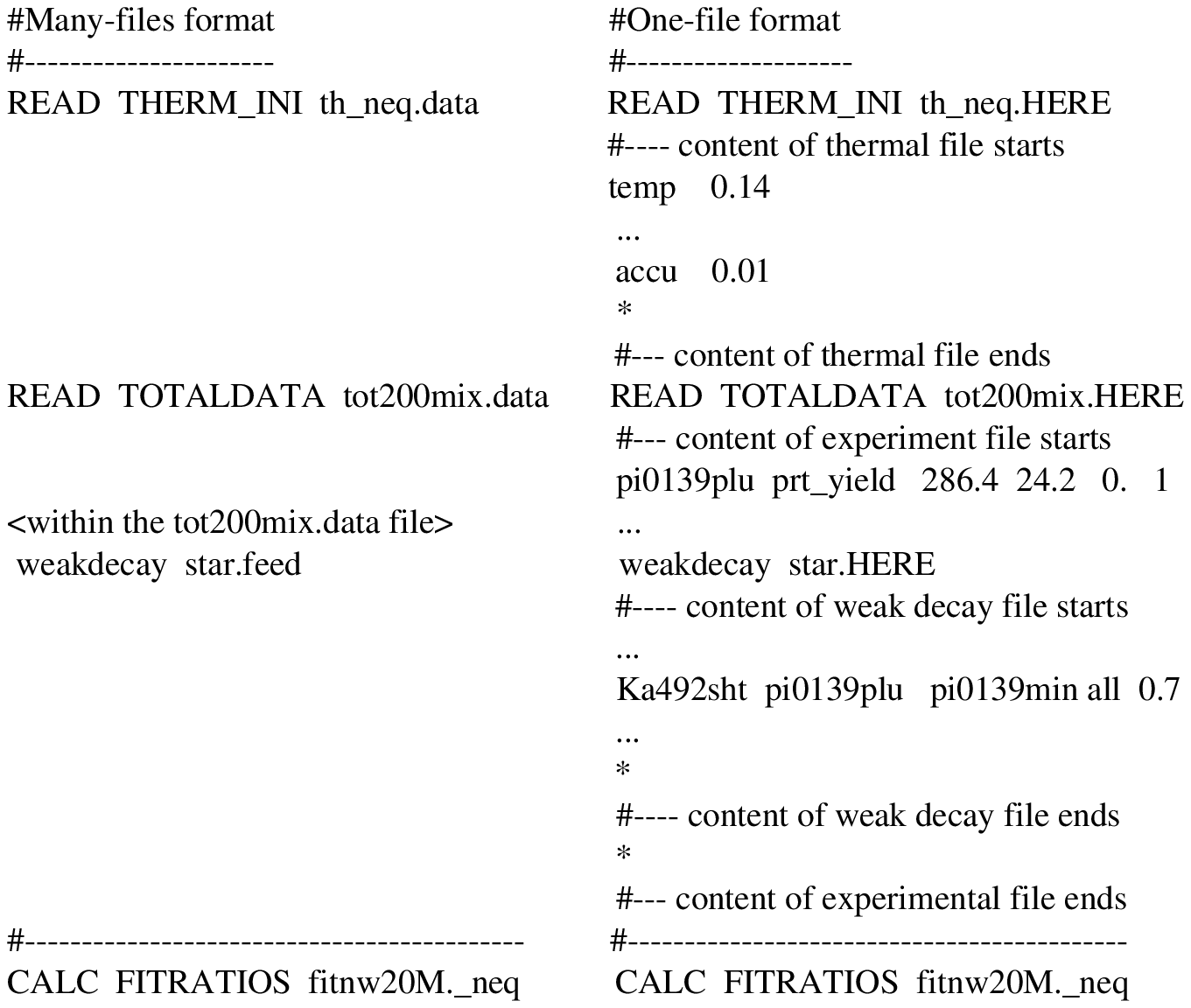}
 \caption{Left:  sharerun.data calling other input files.  Right: One-file format. \label{here}}
\end{figure}

See Fig \ref{here}, and the provided file {\bf sharerun.data$\_$onefile}, for an example 
of how this works.  



\subsection{Combining data points \label{combo}}
SHAREv2 gives the possibility to refer to a different data point within the given fit, and/or combine 
two data points, in order to fit the sum or a product of two particles.  
This feature was described, in the case of fluctuations of ratios, in section \ref{fluctshare} of this paper.

The referring data point consists of one or two (for a combination) numbers, corresponding to the position, 
in the input file, of the point(s) being referred to.

Two numbers united by an operation sign ($ +,\ -$, X, /) will 
add, subtract, multiply and divide two data points.
For instance, if the first data points from the top of the file (see \cite{share_orig} 
section 3.4, for a detailed explanation of the format) are:

\begin{tabular}{llllll}
{\bf   Lm1115zer}  & {\bf  pi0139min}  &  {\bf   Data} & {\bf   $\Delta_{\rm stat}$} &
                                                      {\bf   $\Delta_{\rm syst}$} & {\bf  Fit?}\\
{\bf   Lm1115zrb } & {\bf  pi0139plu }  & {\bf   Data} & {\bf   $\Delta_{\rm stat}$} &
                                                      {\bf   $\Delta_{\rm syst}$} & {\bf  Fit?}\\
\end{tabular}

\noindent then,

\begin{tabular}{llllll}
{\bf   01X02$\phantom{XXX.}$}  &   {\bf  prt$\_$yield } &  {\bf   Data}&  
         {\bf   $\Delta_{\rm stat}$}& {\bf   $\Delta_{\rm syst}$}& {\bf  Fit?}\\
\end{tabular}

\noindent will fit $(\Lambda \overline{\Lambda})/(\pi^+ \pi^-)$, while

\begin{tabular}{llllll}
{\bf   01$\phantom{.XXXXX}$} &   {\bf  fluct$\_$dyn}  &  {\bf   Data}& 
                  {\bf    $\Delta_{\rm stat}$} & {\bf   $\Delta_{\rm syst}$}&  {\bf  Fit?}\\
\end{tabular}

\noindent will fit the dynamical $\Lambda/\pi^-$ fluctuation, as described in section \ref{fluctshare}.

\noindent To fit $(\Lambda+ \overline{\Lambda})/(\pi^+ + \pi^-)$ (but NOT the separate yields), 
the input file will read:

\begin{tabular}{llllll}
{\bf   Lm1115zer} &   {\bf  prt$\_$yield } &   {\bf   Data} &  {\bf   $\Delta_{\rm stat}$}&    
{\bf    $\Delta_{\rm syst}$}& {\bf  
0}\\
{\bf   Lm1115zrb} &   {\bf  prt$\_$yield} &    {\bf   Data}&   {\bf   $\Delta_{\rm stat}$}&    
{\bf $\Delta_{\rm syst}$}&   {\bf 
0}\\
{\bf   pi0139plu } &  {\bf  prt$\_$yield }&    {\bf   Data} &  {\bf   $\Delta_{\rm stat}$}&   
{\bf  $\Delta_{\rm syst}$}&   {\bf 
0}\\
{\bf   pi0139min } &  {\bf  prt$\_$yield} &    {\bf   Data} &  {\bf   $\Delta_{\rm stat}$}&   
{\bf  $\Delta_{\rm syst}$ }&  {\bf 
0}\\
{\bf   01+02$\phantom{XXX.}$} &   {\bf  03+04}  &         {\bf Data} &  {\bf   $\Delta_{\rm stat}$} & 
{\bf  $\Delta_{\rm
syst}$}&{\bf   1}
\end{tabular}

\hspace*{0.1cm}

\noindent {\bf NOTE:} SHARE was written in FORTRAN77.  Feature mentioned in this subsection use 
implicitly recursive code.   SHAREv2 has been tested on several compilers 
and platforms, and found to work. However, compilers and operating systems vary 
--- we would like to know if and when you 
experience problems.

\subsection{Miscellaneous}
The following (small) modifications were made in SHAREv2 compared to SHAREv1:
\subsubsection{Expanded parameter set}
The expanded parameter set includes as noted before, Eq.\,(\ref{gam3eq}), 
{\bf gam3} which allows to 
incorporate a different $u,d$-flavor phase space occupancy. 
A further new variable {\bf dvol}
 describes statistical pressure ensemble fluctuations in volume
(Section 1, Eq.\,(\ref{volfluctdef})).  The provided input
file sets and fixes {\bf dvol} to zero and {\bf gam3} to unity, since experimental measurements sensitive
to these parameters have not as yet been published.  

All details about how to configure these parameters, and fix or relax
them in the context of fits to experimental data, are unchanged w.r.t.
v1.x, described in \cite{share_orig} sections 3.1, 3.6 and 3.7

\subsubsection{Data point sensitivity analysis}
Command {\bf DFIT}, within the file {\bf sharerun.data}  can turn on and 
off the given data point as a point to be fitted.  \\
The syntax for this command is\\
{\bf DFIT  [Datapoint n.]  [Fit(0/1)]}\\
 where {\bf Datapoint n.} refers to the data point's position in the experimental data file from the top, while
{\bf Fit(0/1)} turns this point on (1) or off (0) as a point to be fitted.
For instance, the following input in {\bf sharerun.data}:\\
{\bf READ  TOTALDATA  tot200mix.data}\\
{\bf DFIT  5   1}\\
{\bf CALC  FITRATIOS  fitnw20M-kpi}\\
{\bf DFIT  5   0}\\
{\bf CALC  FITRATIOS  fitnw20M-nkpi}\\
performs two fits.\\
 The first, saved in file {\bf fitnw20M-kpi}  uses the 5th data point 
in {\bf tot200mix.data} when calculating the $\chi^2$ (to be
minimized).\\
The second one, saved in file  {\bf fitnw20M-nkpi}, does not.

\subsubsection{Data point sensitivity profiles}
 Command {\bf SNSPROFIL} calculates the data point {\em sensitivity}.  
The sensitivity is defined as the ratio between the data
point's SHM prediction for a given statistical parameter, and SHM 
prediction at the {\em best fit} value for that parameter.

The syntax of {\bf SNSPROFIL} is the same as {\bf DATPROFIL} in 
\cite{share_orig}, section 4.  The two commands operate in the same way:  
all parameters, except the one on the abscissa, are
minimized at each point in the profile.\\
Thus, the command\\
{\bf CALC SNSPROFIL  temp  0.1   0.2   100  5}\\
will calculate a sensitivity profile for the temperature, going from 0.1 to 0.2
GeV, with 100 points, of the fifth data point within the experimental data-file. 

\subsubsection{Additional output in $\chi^2$ and statistical significance profiles}
 $\chi^2$ profile commands now output the following files:
\begin{description}
\item[name.log] A fit output for each point in the $\chi^2$ profile, in the same 
format as the usual fit output file
(\cite{share_orig}, Fig. 4).
\item[name.chi2, name.stsg] Commands {\bf SNSPROFIL} and {\bf DATPROFIL} also output 
the $\chi^2$ profile (extension *.chi2) and
$P_{true}$ profile (extension *.stsg). 
\end{description}
\subsubsection{Improved treatment of fit errors}
SHAREv2 automatically runs the MINOS algorithm \cite{minuit} if the fit
can not get a robust estimate of the errors.   This results in a considerable
improvement of error-handling.  This update entails no changes in the user interface or output format.

\section{Comparison with previous versions}
\subsection{Testing SHAREv2}

SHAREv2 was extensively tested for programming and physics errors:
\begin{itemize}
\item SHM Calculations and fit results for SPS and RHIC energies were verified to be equal between
SHAREv2 and SHAREv1.x reference results;
\item SHAREv2 reads SHAREv1.x weak decay input.  The equivalence between the 
two treatments, when weak decay files are designed to reproduce SHAREv1.x format, 
was shown to all decimal places;
\item Fluctuations of conserved quantities (such as $\ave{(\Delta Q)^2}$) were 
compared before and after resonance decays.   The conservation of this quantity 
implies that the enhanced fluctuations after all resonances decayed are exactly
 balanced by multiplicity correlations between the resonance decay products.  
This holds true to two decimal places (up to two-step correlations
arising from decays such as $K(1600) \rightarrow K(892) \pi \rightarrow K \pi \pi$.  
These correlations are not tracked by SHARE, but their contribution is below $1 \%$).
\end{itemize} 

\subsection{SHAREv1.x bugs found}
While developing and testing SHAREv2,
several minor bugs and choice issues were found in the previous version SHAREv1.x,
 The most noteworthy issues which lead to sometimes noticeable 
(beyond line width) changes in the results  are:
\begin{description}
\item[SHAREv1.1, v1.2] The Bessel function series was incorrectly truncated for
large $\gamma_q$ (close to pion B--E condensation);
\item[SHAREv1.3] Quark flavor mixing   error  in calculation of
mesons such as $\eta$ and $\phi$, for $\gamma_q \ne \gamma_s$;
\item[SHAREv1.1--v1.3] The most relevant issue is actually 
 not an error but lack of versatility in
the handling of $\Sigma \rightarrow p \pi$ decays:  $\Sigma$-particles
decay weakly, like the $\Lambda$s and the $\Xi$s.   However, unlike 
$\Lambda$ and $\Xi$,   $\Sigma$-decays
are not experimentally reconstructible since at least one of 
  $\Sigma$-decay products is neutral. In general, 
SHAREv1.x particles from these decays were included in the yield 
count. However, it turned out that while some experiments had much less 
than full acceptance for these decays,   other  experiments, e.g., NA49, have removed
 $\Sigma \rightarrow p$ feed-down via Monte-Carlo simulations accounting 
for the experimental acceptance of the decay products, with $\Sigma$ 
yields obtained from the  observed $\Lambda$-yields. 
    \end{description}
 
Working with patched SHAREv1.x,
we realized that $\Sigma$-decay  issue mattered in  that some fits got better 
allowing for a modified $\Sigma$ weak decay pattern.
Issues, such as this one, prompted us to introduce a more general treatment 
of weak decay particle yield contributions in SHAREv2. 

\section{Installation} The\footnote{This chapter is nearly identical to the
 {\bf READMEv2.html} file found on the SHARE webpage}
 SHAREv.2.1 program code and input files are contained
in a tar.gz archive.
The file  sharev2.1.tar.gz is available at\\
{\tt http://www.physics.arizona.edu/$\tilde{\phantom{~}}$torrieri/SHARE/share.html }\\
To unpack it, create a SHARE directory, put the archive
in it, and execute the following commands:

\noindent {\bf gunzip  sharev2.1.tar}\\
{\bf tar -xf  sharev2.1.tar}

The following files will then be created, enough for a
complete `representative' run of SHARE:
\begin{description}
  \item[decays.data]     The complete Particle Data Group decay tree
                  (section 3.3 in \cite{share_orig});
  \item[dec$\_$no.data]     An empty decays file, useful for testing  the program
                  calculations (abundancies reduce to modified Bessel functions)
                  as well as studying the role of resonances in stable particle 
                  ratios;
  \item[fortrat]         A shell script compiling (in f77) the FORTRAN code
                  which should be modified depending on location of FORTRAN
                  (g77 or f77 or f95) and CERN library of programs;
  \item[particles.data]  Particle properties, with full widths
                  (section 3.2, in \cite{share_orig}));
  \item[partnowdt.data]  Particle properties, with no widths.
                  Calculations with this input file require
                  considerably less computational time, and it
                  suffices when there are no resonances in the fit;
  \item[ratioset.data]   The FORTRAN fit input file
                  (section 3.5, in \cite{share_orig});
  \item[sharerun.data]   A `representative' run input file
                   (section 4, in \cite{share_orig});
                  including an analysis of fluctuations and yields similar to what 
                  was presented by members of our collaboration in \cite{mefluct2};
  \item[sharerun.data$\_$onefile]   The same file in the single file format, as explained 
in section \ref{singlefile};
  \item[samplefit200]    A directory containing the output files generated 
by running the provided `sharerun.data',
                  as a debugging/comparison standard;
  \item[sharev2.1.f]     SHAREv1.1 FORTRAN code.    The header contains information about bug fixes;
  \item[thermo.data]     A representative thermal parameter input file
                  (section 3.1,  in \cite{share_orig}).
                  It is set to reasonable non-equilibrium fit values;
  \item[totbar200.data]  A representative data input file
                  (section 3.4,  in \cite{share_orig})
                  containing ratios, yields and fluctuations drawn from RHIC experiments, as of July 2004 
                  (see references in \cite{mefluct2});
   \item[star.feed]      An example of a decay feed-down coefficients file.
                  See section \ref{weakdecaysect}.
\end{description}

Note that SHARE requires CERN libraries, to be downloaded separately 
from\\ \textit{http://wwwasd.web.cern.ch/wwwasd/cernlib.html}

The compiler statement (in file `fortrat') is\\
\textit{f77  -L/usr/local/cern/pro/lib -o sharev2.1.exe sharev2.1.f  -lmathlib -lkernlib -lpacklib -C}\\
which assumes that the CERN libraries are in directory \\
\textit{/usr/local/cern/pro/lib}\\
If this is not true on your system, fortrat should be changed accordingly.  
                 
Once the directory is unpacked, the program should be compiled with

\noindent \textbf{./fortrat}

After this, typing

\noindent \textbf{./sharev2.1.exe}
\\should produce a correct run with a detailed output which
shows the program's capability.    A copy of these files, produced on our computer system, is also included in {\bf sharev2.1.tar.gz} within a directory called {\bf samplefit200} for comparison with the files produced by the installed program. 
Several output files are produced, with the following names
as default.
The contents of each file are explained in detail in \cite{share_orig} section 4.
\begin{description}
\item[fit*.out]     Fit output files;
\item[graph*]       Fit output graphics (experiment, fitted values, calculated values);
\item[prof*]        $\chi^2$ profiles and correlation functions for the various fits.
\end{description}
See \cite{share_orig} about more details about these files's contents.
\section{Status, conclusions, and future plans}
One of the areas of current intense interest in the field of high energy heavy ion 
reactions is the understanding of the mechanisms of soft hadron production
(chemical freeze-out), that is
the study of how the energy confined in the central fireball turns into
matter in a multi particle production process.  

The SHARE suite of programs is an analysis tool of particle yields addressing    
 the following questions:
\begin{itemize}
\item What is the chemical freeze-out temperature, chemical potentials and volume?
\item What are the physical properties of the fireball which hadronizes?
\item Is the hadron system in chemical equilibrium at freeze-out?
\item What is the degree of re-interaction between hadronization and freeze-out?
\end{itemize}

The need for SHARE arises from recognition that the book keeping task involved in the correct
application of the statistical hadronization model is considerable, often transcending the 
resources available to individual researchers. The current SHAREv2 program follows 
on SHAREv1.x \cite{share_orig}
adding three significant novel features: 
a) flexible handling of particle decay feed-down, 
b) fluctuations,  and 
c) complete $u,\,d,\,s,\,c$  flavor content treatment. We note that, since SHAREv1.x was released, 
another analysis package appeared,   THERMUS \cite{thermus}. THERMUS has 
more convenient setup for the experimental data analysis environment. On the other
hand, THERMUS is  not  addressing many 
of the features SHARE offers, including  chemical non-equilibrium and now 
in the current SHAREv2, fluctuations, and light flavor details. 

SHARE allows an analysis of experimental data  that can 
address indirectly questions related to the dynamics of the fireball 
evolution, since  chemical non-equilibrium implies  a fast hadronization.
However, the analysis of particle yields  cannot give results of greater 
precision than is inherent in the data it treats, and this is not 
yet good enough to resolve the question about the nature of SHM as seen in the recent
references \cite{equil_energy,gammaq_energy,gammaq_size,gammaq_hyperons,cleymans2}, and the
possible relation to the onset of a phase transition, \cite{hornfuture}.
SHAREv2 offers the additional analysis feature, the fluctuation in hadron yields and ratios,
which will help to settle  these issues, should  precise particle yield 
experimental data  not become available. Moreover, SHAREv2 through the study 
of the consistency of particle yield and fluctuation, can test the  SHM
in depth.

The development of phenomenological tools capable of falsifying statistical 
hadronization models 
is, of course, far from over.   Possible extensions of chemical freeze-out model, in
 future version of SHARE, might  include a canonical ensemble module, allowing to test 
SHM chemical non-equilibrium in small physical systems, and  the introduction
of an opacity parameter to correct the yields of observed  resonances \cite{resonanceAB}.
Another possible future development would entail extending SHARE 
towards a detailed description of momentum 
distributions. This would be somewhat different from THERMINATOR~\cite{therminator}
model which
relies on scaling in rapidity, and thus, only applies to ultra high energy collisions. 
We note that particle
momentum distributions are dependent, in addition to the physics 
incorporated in SHAREv2 also
on the dynamical  evolution of  the emitting source, and on the degree of resonance 
rescattering following on the chemical freeze-out. Strategies how these two 
interwoven effects can be disentangled are being developed. For further discussion 
of these and related questions we refer to a recent review~\cite{Florkowski}.  


\end{document}